\documentclass[preprint,sort&compress]{elsarticle}

\usepackage{mathtools}
\usepackage{amssymb}
\usepackage{url}
\usepackage{booktabs}
\graphicspath{{./figures/}}

\DeclareMathOperator{\diag}{diag}
\DeclareMathAlphabet{\mathbf}{OML}{cmm}{b}{it}
\newcommand{\eps}{\varepsilon}
\newcommand{\T}{\mathcal{T}}
\renewcommand{\i}{\mathrm{i}}
\renewcommand{\vec}[1]{\mathbf{#1}}

\journal{Computer Physics Communications}

\begin{document}

\begin{frontmatter}

\title{Imaginary time propagation code for large-scale two-dimensional
eigenvalue problems in magnetic fields}

\author[nsc]{P.J.J.~Luukko\corref{cor}}
\ead{perttu.luukko@iki.fi}

\author[tut,nsc]{E.~R\"as\"anen}

\address[nsc]{Nanoscience Center, University of Jyv\"askyl\"a, PL 35, FI-40014, Finland}

\address[tut]{Department of Physics, Tampere University of Technology, FI-33101
Tampere, Finland}

\cortext[cor]{Corresponding author}

\begin{abstract}

	We present a code for solving the single-particle, time-independent
	Schr\"odinger equation in two dimensions. Our program utilizes the
	imaginary time propagation (ITP) algorithm, and it includes the most recent
	developments in the ITP method: the arbitrary order operator factorization
	and the exact inclusion of a (possibly very strong) magnetic field. Our program
	is able to solve thousands of eigenstates of a two-dimensional quantum
	system in reasonable time with commonly available hardware. The main
	motivation behind our work is to allow the study of highly excited states
	and energy spectra of two-dimensional quantum dots and billiard systems with
	a single versatile code, e.g., in quantum chaos research. In our
	implementation we emphasize a modern and easily extensible design, simple and
	user-friendly interfaces, and an open-source
	development philosophy.

\end{abstract}


\end{frontmatter}


\noindent\textbf{Program summary}

{\small\noindent
\emph{Program title:} itp2d\\
\emph{Journal reference:}\\ 
\emph{Catalogue identifier:}\\ 
\emph{Licensing provisions:} GNU General Public License, version 3\\
\emph{Programming languages:} C++ and Python\\
\emph{Computer:} Tested on x86 and x86-64 architectures.\\
\emph{Operating system:} Tested under Linux with the g++ compiler. Any POSIX-compliant OS with a C++ compiler and the required external routines should suffice.\\
\emph{RAM:} 1 MB or more, depending on system size.\\
\emph{Has the code been vectorised or parallelized?:} Yes, with OpenMP.\\
\emph{Classification:} 7.3\\
\emph{External routines/libraries:} FFTW3 (\url{http://www.fftw.org}), CBLAS (\url{http://netlib.org/blas}), LAPACK (\url{http://www.netlib.org/lapack}), HDF5 (\url{http://www.hdfgroup.org/HDF5}), OpenMP (\url{http://openmp.org}), TCLAP (\url{http://tclap.sourceforge.net}), Python (\url{http://python.org}), Google Test (\url{http://code.google.com/p/googletest/})\\
\emph{Nature of problem:} Numerical calculation of the lowest energy solutions
(up to a few thousand, depending on available memory), of a single-particle,
time-independent Schr\"odinger equation in two dimensions with or without a
homogeneous magnetic field.\\
\emph{Solution method:} Imaginary time propagation (also known as the diffusion
algorithm), with arbitrary even order factorization of the imaginary time
evolution operator.\\
\emph{Additional comments:} Please see the README file distributed with the program for more information. The source code of our program is also available at \url{https://bitbucket.org/luukko/itp2d}.\\
\emph{Running time:} Seconds to hours, depending on system size.
}


\section{Introduction}
	In this paper we present \texttt{itp2d}: a modern implementation of
	the imaginary time propagation (ITP) scheme for solving the eigenstates of
	a two-dimensional, single-particle quantum system from the time-independent
	Schr\"odinger equation. Our implementation includes the most recent
	developments in imaginary time propagation, namely, the arbitrary order
	operator factorization by Chin~\cite{chin}, and the exact method
	of including a homogeneous magnetic field developed by
	Aichinger~\textit{et~al.}~\cite{itpmagn,gauge}.

	The computational methods used in our program have been implemented
	before~\cite{ndsch,itpmagn,olderitp}, but \texttt{itp2d} is the first to
	combine them all in a two-dimensional code. In addition, with
	\texttt{itp2d} we focus on a clear, object-oriented, and extensible
	implementation, without compromising efficiency. We have also emphasized on
	making the implementation suitable for solving a great number of
	eigenstates -- up to thousands -- in reasonable time. With this we intend
	to make \texttt{itp2d} a useful tool for studying the energy level
	spectra and highly excited wave functions of two-dimensional quantum
	systems, for example in the field of quantum chaos~\cite{stockmann}. Since solving the
	single-particle equation is also a crucial part in the many-particle
	formalism of density-functional theory, \texttt{itp2d} can also be used as
	a fast eigensolver for realistic electronic structure calculations.

	There are other algorithms designed for probing the highly excited states
	of quantum billiards, such as the various boundary methods~(see, e.g.
	Refs.~\cite{boundary1,boundary2} and references therein), but we have
	chosen ITP for its versatility and its ability to handle strong magnetic
	fields. To our knowledge, ITP with its latest developments has not been
	thoroughly benchmarked against other general-purpose iterative
	eigensolvers. In one reported case~\cite{lehtovaara}, ITP was found to have
	a much better scalability with respect to the grid size compared to the
	Lanczos method, but worse scalability with respect to the number of states.
	However, these conclusions were based on a single 3D system solved only up
	to 10 eigenstates with the older, 4th order operator factorization. In
	Sec.~\ref{s:benchmarks} we present benchmark results which indicate that
	\texttt{itp2d} provides competitive numerical efficiency compared to
	publicly available, general-purpose eigensolvers.

	ITP can and has been~\cite{ndsch} implemented in any number of dimensions.
	We have chosen to limit our implementation to two dimensions for
	simplicity, and because operating in two dimensions allows systems with
	thousands of eigenstates to fit into system RAM, minimizing slow I/O
	operations.

	In order to give a reasonably self-contained presentation, we will first
	introduce the ITP scheme and the underlying algorithms used in our
	implementation. A reader who is already familiar with ITP can advance
	directly to Sec.~\ref{s:implementation}, where we will describe the
	implementation in more detail.

	All formulas in this paper are given in Hartree atomic units. The
	magnetism-related units follow the SI-based convention, i.e., the atomic
	unit of magnetic field is $\hbar/ea_0^2$, where $e$ is the unit charge and
	$a_0$ is the Bohr radius.

\section{Algorithms}

\subsection{Imaginary time propagation}
	\label{s:itp}
	Imaginary time propagation (also known as the \emph{diffusion method}) is a
	general algorithm for solving eigenvalue problems such as the
	time-independent Schr\"odinger equation
	\[
		H\psi = E\psi,
	\]
	where~$H$ is the Hamiltonian of the system, and the solution of the
	eigenvalue problem is some set of energies~$\{E_j\}$ and the associated
	eigenstates~$\{\psi_j\}$.
	
	The ITP algorithm is based on the idea that the (initially unknown)
	eigenstates of~$H$ form a complete basis of the associated Hilbert space,
	and thus any arbitrary state~$\phi$ can be expanded in the eigenstates
	of~$H$ as
	\[
		\phi = \sum_j c_j\,\psi_j,
	\]
	where $\{c_j\}$ is some set of coefficients. By choosing an operator $\T =
	\exp(-\eps H)$, where $\eps > 0$, we can ``filter out'' the lower energy
	states out of this expansion by repeatedly applying~$\T$ to our initial
	state~$\phi$. After one iteration we get
	\begin{equation}
		\label{eq:itp}
		\T\phi = \sum_j c_j\,\T\psi_j = \sum_j (c_j\exp(-\eps E_j)\,\psi_j,
	\end{equation}
	in other words, components in the expansion are damped with the damping
	coefficient approaching zero exponentially with increasing energy and
	increasing value for~$\eps$. By repeatedly applying the operator~$\T$ and
	subsequently normalizing the state, all higher energy components are
	removed from the expansion and only the ground state of the Hamiltonian
	remains. The operator~$\T$ has the same form as the time evolution operator
	of a quantum system in the Schr\"odinger picture, $\exp(-\i tH)$, only with
	an imaginary number inserted for time, giving the ITP algorithm its name.
	For this reason the value~$\eps$ is also called the \emph{imaginary time
	step}.

	In order to get more states besides the ground state one chooses a set of
	arbitrary, linearly independent initial states and repeatedly applies~$\T$
	on each state followed by orthonormalizing the set of states. By choosing a
	set of~$N$ initial states this procedure leads to the set of~$N$ lowest
	energy eigenstates of the Hamiltonian. The convergence speed of each state
	typically depends on its energy, with the ground state converging first and
	the highest energy states last. This also means that often it is
	advantageous to include some extra states in the computation -- with more
	states each iteration takes longer, but since the absolutely highest states
	are not required to converge, the number of required iterations is reduced.

	Since the method is based on removing unwanted components to project
	out the wanted ones, it is essential that the initial states really
	include all desired components in the first place, i.e., even though the
	initial states are otherwise arbitrary, all of them should not be
	orthogonal to any eigenstate. A simple way to make sure that this
	requirement is practically always fulfilled is to choose random noise as
	the initial states.

	To apply the ITP method in computer simulations we need a method to
	compute~$\T\phi$ for an arbitrary state~$\phi$, an efficient method to
	orthonormalize a set of states, and a method to check for convergence, i.e.
	to assess whether the current set of states is close enough to the true
	eigenstates of~$H$. After these steps are implemented the iterative ITP
	scheme can be used as summarized in the following:
	\begin{enumerate}
		\item Start with a set of~$N$ arbitrary, linearly independent
			states~$\phi_n$.
		\item Apply the operator~$\T = \exp(-\eps H)$ to each
			state\label{it:startiter}.
		\item Orthonormalize the set of states.
		\item Check whether convergence has been achieved. If yes, terminate,
			if not, go back to step~\ref{it:startiter}.
	\end{enumerate}

\subsection{Orthonormalization}
	\label{s:ortho}
	The orthonormalization of states can be executed with a standard
	Gram-Schmidt process, but in our implementation we have opted for the
	subspace orthonormalization algorithm~\cite{lowdinreview}, which has been shown to be better suited for ITP in
	previous implementations~\cite{olderitp}. In the Gram-Schmidt method one
	state is chosen to be a starting point of the iterative orthonormalization
	scheme, whereas the subspace orthonormalization method treats all states equally.

	The subspace method for orthonormalizing a set~$\{\phi_j\}$ of $N$ linearly
	independent states can be summarized as follows. First the overlap
	matrix~$M$ with $M_{ij} = \langle\phi_i|\phi_j\rangle$ is computed. Then
	the Hermitian matrix~$M$ is diagonalized, which results in a unitary
	matrix~$U$ and a diagonal matrix~$D = \diag(m_1, m_2, \dotsc, m_N)$ such
	that $M = UDU^\dagger$. Now the linear combinations
	\begin{equation}
		\label{eq:lincomb}
		\phi_i' = \frac{1}{\sqrt{m_i}}\sum_j U_{ji}\phi_j
	\end{equation}
	form an orthonormalized set, which can be confirmed easily by a direct
	calculation of the inner product $\langle\phi_i'|\phi_j'\rangle$. In total,
	the subspace orthonormalization algorithm for~$N$ states amounts computationally to
	\begin{enumerate}
		\item Computing $N(N+1)/2$ inner products to form the overlap matrix~$M$
		\item Diagonalizing a $N\times N$ Hermitian matrix~$M$
		\item Forming the linear combinations~\eqref{eq:lincomb} which, if the
			states are stored as~$N$ arrays of $M$ numbers each, amounts to the
			computation of a matrix product between a~$N\times N$ unitary
			matrix and a $N\times M$ matrix
	\end{enumerate}
	All these steps can be implemented easily and efficiently with standard
	linear algebra routines. In addition, all these steps can be computed
	without allocating memory for a second set of states.

	The subspace orthonormalization also provides a way to approximate the energy
	of each state based on the eigenvalues $\{m_n\}$, since with successive
	iterations $m_n \rightarrow \exp(-2\eps E_n)$. However, in systems with a
	high number of states, the energy values~$E_n$ become very large, which causes
	the values for~$m_n$ get very close to zero. In turn, the accuracy
	of this approximation becomes quite poor.

	Regardless of the algorithm used, a requirement for orthonormalization to
	work is of course that the states are linearly independent. This can be
	asserted for the initial states, but it can happen that the states lose
	linear independence during the propagation. If the states are propagated
	with a too large time step several states can, for example, get so close to
	the ground state that they are essentially linearly dependent. This poses
	limitations on how large time steps can be used.

\subsection{Factorization of the imaginary time evolution operator}
	For practical Hamiltonians the exponential $\T = \exp(-\eps H)$ can not be
	implemented directly in computations. The traditional approach to this
	problem is to approximately factorize the exponential into an easier form.
	For a Hamiltonian of the form $H = T + V$, the most simple approximation is
	the second-order factorization
	\begin{equation}
		\label{eq:T2}
		\T = \exp(-\tfrac{1}{2}\eps V)\,\exp(-\eps T)\,\exp(-\tfrac{1}{2}\eps
		V) + \mathcal{O}(\eps^3).
	\end{equation} If $T$ represents the kinetic energy operator, $T \propto
	\nabla^2$, and $V$ is a local potential operator, both remaining
	exponentials $\exp(-\tfrac{1}{2}\eps V)$ and $\exp(-\eps T)$ can be
	implemented easily: the exponential of~$V$ is still a local potential, so
	in the coordinate basis it is diagonal, and likewise the exponential of~$T$ is
	diagonal in the wave vector basis. This means that for wave functions in the
	coordinate basis an exponential of~$V$ is simply a pointwise
	multiplication, and an exponential of~$T$ is a combination of a Fourier
	transform and a pointwise multiplication.

	As pointed out, the factorization in Eq.~\eqref{eq:T2} is only approximate
	for general Hamiltonians. When using an approximate evolution
	operator~$\T'$, we essentially replace the original Hamiltonian~$H$ with an
	approximation~$H'$ such that $\T' = \exp(-\eps H')$, and the eigenstates we
	get with the ITP method are actually the eigenstates of~$H'$. The better
	the approximation for~$\T$, the more accurately the eigenenergies and
	eigenstates of $H'$ match the true result we would get for~$H$. Since some
	kind of approximation for~$\T$ is required we need to find a balance
	between two opposing effects when choosing a value for the imaginary time
	step~$\eps$: a larger value for~$\eps$ causes the ITP scheme to converge
	\emph{faster} (as seen directly from the ``damping factor'' in
	Eq.~\eqref{eq:itp}), but the $\eps$-dependent approximation causes the
	scheme to converge \emph{further away} from the true solution.

	In order to improve the ITP method several improved, higher order
	approximations for~$\T$ have developed beyond the second-order
	factorization of equation~\eqref{eq:T2}. The most recent improvement is the
	factorization by Chin~\cite{chin}, that constructs an \emph{arbitrarily
	high order} approximation for~$\T$ from the second-order factorization:
	\begin{equation}
		\label{eq:chin}
		\T = \sum_{k=1}^n c_k\,\T_2^k(\eps/k) + \mathcal{O}(\eps^{2n+1}),
	\end{equation}
	where $\T_2(\eps)$ is the second-order approximation from Eq.~\eqref{eq:T2},
	and the coefficients~$\{c_i\}$ are given by
	\[
	c_i = \prod^n_{\mathclap{\substack{j = 1\\j \neq i}}} \frac{i^2}{i^2 - j^2}.
	\]

	Using a higher-order approximation for~$\T$ allows for higher values
	of~$\eps$, which improves the convergence rate of the ITP scheme. This is
	highly advantageous, since even though more complicated approximations
	for~$\T$ make the propagation step more computationally intensive, fewer
	iterations are needed due to the faster convergence rate. As the number
	of states~$N$ is increased, the computational cost of the propagation step
	in the ITP scheme scales as~$\propto N$ (each state is simply propagated
	independently of the others), but the orthonormalization step usually
	scales as~$\propto N^2$ or worse. This means that regardless of how
	complicated the propagation operator~$\T$ is, the orthonormalization step
	starts to quickly dominate the computation completely, and thus for solving
	a high number of states it is critical to keep the total number of
	iterations at a minimum by using a high-order approximation for~$\T$.
	However, since higher-order factorizations involve an increasing number of
	arithmetic operations, finite precision arithmetic poses limits on how high
	order expansions of type~\eqref{eq:chin} are reasonable. As reported in the case of a
	separate implementation of ITP \cite{ndsch}, we confirm that order~12 is
	usually the limit for double-precision arithmetic.

\subsection{Including a homogeneous magnetic field}
	\label{s:magnetic}
	In the presence of a magnetic field~$\vec{B}$ characterized by a vector
	potential~$\vec{A}$, the canonical momentum operator of an electron is, in
	SI-based Hartree atomic units,
	\[
		\Pi = -\i\nabla + \vec{A},
	\]
	and the kinetic energy operator is $T = \tfrac{1}{2}\Pi^2$. This operator
	is no longer diagonal in wave vector space, so applying the
	operator~$\exp(-\eps T)$ is no longer trivial. However, as noted by
	Aichinger~\textit{et~al.}~\cite{itpmagn}, when the magnetic field is homogeneous and
	parallel to the $z$-axis, the required exponential term can be
	factorized~\emph{exactly}:
	\begin{multline}
		\label{eq:magnfact}
		\exp(-\eps T) = \exp(-\tfrac{\eps}{2}(\Pi_x^2 + \Pi_y^2 + \Pi_z^2))\\=
		\exp(-\tfrac{\eps}{2}f_x(\xi)\Pi_x^2)\,
		\exp(-\tfrac{\eps}{2}f_y(\xi)\Pi_y^2)\,
		\exp(-\tfrac{\eps}{2}f_x(\xi)\Pi_x^2)\,
		\exp(-\tfrac{\eps}{2}\Pi_z^2),
	\end{multline}
	where $\Pi_x$, $\Pi_y$ and $\Pi_z$ are the $x$-, $y$- and $z$-components of
	$\Pi$, respectively, and the coefficients~$f_x$ and $f_y$ are
	\begin{equation}
		\label{eq:fxfy}
		f_x(\xi) = \frac{\cosh(\xi)-1}{\xi\sinh(\xi)}\quad\text{and}\quad
		f_y(\xi) = \frac{\sinh(\xi)}{\xi},
	\end{equation}
	given as a function of $\xi = \epsilon B$.

	The next step is choosing the gauge of the vector potential in a way that
	each of term in factorization~\eqref{eq:magnfact} can be
	implemented efficiently. The linear gauge~$\vec{A} = (-By, 0, 0)$, where
	$B$ is the magnetic field strength, is a good choice, because then the components
	of~$\Pi$ are simply $\Pi_x = k_x - By$, $\Pi_y = k_y$, $\Pi_z = k_z$, in
	terms of the wave vector~$\vec{k} = (k_x, k_y, k_z)$. This
	means that the factorized~$\exp(-\eps T)$ can be applied to a wave function
	by first Fourier transforming from the $(x, y, z)$ basis to $(k_x, y, k_z)$,
	where both $\exp(-\tfrac{\eps}{2}\Pi_z^2)$ and
	$\exp(-\tfrac{\eps}{2}f_x(\xi)\Pi_x^2)$ are diagonal and easily applied.
	Then we can Fourier transform the remaining $y$-coordinate in order to get to
	the basis $(k_x, k_y, k_z)$ where $\exp(-\tfrac{\eps}{2}f_y(\xi)\Pi_y^2)$ is
	diagonal. Finally we transform back to $(k_x, y, k_z)$ in order to apply
	$\exp(-\tfrac{\eps}{2}f_x(\xi)\Pi_x^2)$ again. All steps require only
	Fourier transforms and pointwise multiplications, and moreover, the
	number of required Fourier transforms is not increased from the case
	of zero magnetic field.

	This method of exact factorization of the kinetic energy part allows for a
	(possibly very strong) homogeneous external magnetic field to be included
	in ITP simulations without any additional approximations. Another
	attractive
	feature of this method is that it can be made gauge-invariant regardless of the
	discretization~\cite{gauge}, removing gauge-origin problems that often affect
	computations with magnetic fields.

\subsection{Treating Dirichlet boundary conditions}
\label{s:dirichlet}
	Throughout the previous discussion, the use of Fourier transforms to go
	from the position to the wave-vector basis has implied the use of periodic
	boundary conditions. Switching to Dirichlet boundary conditions would allow
	the study of billiard systems, which are common model systems in quantum
	chaos research~\cite{stockmann}.

	A simple way to enforce Dirichlet boundary conditions for a rectangular
	calculation box is to replace the Fourier transforms with sine transforms,
	or in other words, to expand the wave functions in eigenstates of a
	particle in a rectangular box instead of plane waves. However, there are
	two complications in this simple approach. First of all, the use of the
	sine transform still implies periodicity across the boundary. The wave
	functions will be periodic because of the Dirichlet boundary conditions,
	but the wave functions can have a discontinuous derivative at the boundary.
	Because of this possible discontinuity of the derivative, the expansions in
	sine waves can have spurious, high-frequency ``ringing'' artifacts. These
	artifacts will be dampened by the ITP iterations, but they will worsen
	convergence.

	Secondly, with an external magnetic field the sine waves are no longer as
	good a basis. For example, applying the Hamiltonian to a combination of
	sine functions results in a combination of sine \emph{and cosine}
	functions, since with a magnetic field the Hamiltonian also includes first
	derivatives. Previously, operators such as the kinetic energy and the
	exponential of kinetic energy turned out to be simple: Fourier transform to
	the wave-vector space, a multiplication, and a transform back. With a
	magnetic field and Dirichlet boundary conditions they become a sine
	transform, \emph{two} multiplications, and \emph{two} inverse transforms,
	because the sine and cosine parts need to be handled separately. The correct
	basis to use would be the eigenfunctions of a particle in a rectangular box
	with a magnetic field, but to this problem no simple solution is known --
	computing these eigenfunctions was a major goal for~\texttt{itp2d}, and the
	reason Dirichlet boundary conditions were implemented in the code.

	It should be pointed out, however, that these problems do not prevent
	combining Dirichlet boundary conditions with an external magnetic field,
	they only cause slower converge. Our implementation can, for example, solve
	the first few thousand eigenstates of the particle in a box with a magnetic
	field. Improving the combination of Dirichlet boundary conditions and a
	magnetic field will be a major goal for future development of the code.

\subsection{Convergence checking}
	\label{s:convergence}
	As discussed previously, the ITP scheme with a fixed imaginary time
	step~$\eps$ converges faster with larger~$\eps$, but to a more inaccurate
	solution. For this reason the ITP scheme is traditionally coupled with time
	step adjustment, i.e., the states are first converged with a larger time
	step and the time step is subsequently decreased, iterating this
	converge-decrease cycle until some final criteria of convergence is
	fulfilled. There are therefore two ``levels'' of convergence involved:
	convergence with respect to the current value of~$\eps$, and final
	convergence.

	A natural measure of final convergence for a state~$\psi$ is the standard
	deviation of energy
	\[
	\sigma_H(\psi) = \sqrt{\langle\psi|H^2|\psi\rangle - \langle\psi|H|\psi\rangle^2},
	\]
	where~$H$ is the system Hamiltonian. This quantity also gives an error
	estimate for the computed eigenenergies. Convergence with respect to the time
	step can be considered by looking at the \emph{change} of $\sigma_H$
	between successive iterations -- when the standard deviation no longer
	decreases by a significant amount between iterations, the state can be
	considered converged with the current time step.

	Simpler measures of convergence can be implemented by looking directly at
	the values of energy obtained at each iteration and considering the state
	converged when either the relative or absolute change in energy between
	successive iterations gets small enough. Another simple way of defining
	final convergence is the point when decreasing the time step seems to be of
	no use, i.e., the point when after decreasing the time step, the state
	converges with respect to the decreased time step with only one iteration.

	Due to the fact that the convergence checks represent in any case an
	insignificant share of the total computational resources, it is usually
	best to use the standard deviation as a measure of convergence. The simpler
	methods come to play only when something prevents the use of the standard
	deviation. This occurs, for example, when using an external magnetic field
	combined with the method of enforcing Dirichlet boundary conditions
	discussed in Sec.~\ref{s:dirichlet}, since the ringing artifacts near the
	edges make the computation of $\langle\psi|H^2|\psi\rangle$ inaccurate.

\section{Implementation}
\label{s:implementation}

\subsection{Program structure overview}

	Our implementation of ITP for two-dimensional systems, \texttt{itp2d}, is
	based on a high-level, object-oriented design, with calls to optimized
	external routines for time-consuming low-level operations. This makes the
	program easier to maintain and extend without compromising computational
	efficiency. All computations are done in SI-based Hartree atomic units and
	for simplicity, Hartree atomic units are also used for all input and output
	(except timing data, which is given in seconds).

	The complete ITP simulation implementation is encapsulated in a single
	high-level C++ class, making our program easily included in separate
	programs needing a fast Schr\"odinger equation solver, e.g., for solving
	the Kohn-Sham equations for density-functional theory calculations. This
	high-level C++ interface is supplemented with a simple (but complete)
	command line interface, that provides an easy way to run simulations with
	different parameters without recompiling. The command line interface is
	implemented using the Templatized C++ Command Line Parser Library (TCLAP).

	The program is distributed with a separate documentation file that covers the
	use of~\texttt{itp2d} from a more practical point of view. The command line
	interface also includes inline documentation, accessible with the command
	line argument \texttt{--help}.

	For compiling~\texttt{itp2d} a simple GNU Makefile is provided. The
	Makefile is designed for the free and portable~\texttt{g++} compiler from
	the GNU Compiler Collection. The actual program code in~\texttt{itp2d}
	should be standards compliant C++, so other standards compliant compilers
	can also be used, but this requires modifications to the Makefile. In a
	similar way, \texttt{itp2d} is only tested on computers running Linux, but
	the program should work in other systems with minimal effort, provided a
	C++ compiler and the required external routines are available.

\subsection{Implementation of the ITP scheme}

	The wave functions operated on by ITP are implemented as two-dimensional
	arrays of double precision complex numbers on a rectangular grid with
	uniform spacing. This low-level memory layout is supplemented with a
	high-level class interface providing the necessary arithmetic operations.
	Similar class interfaces are provided for arrays of wave functions for
	easily handling several wave functions as a whole. Operators acting on
	wave functions are similarly defined in an object-oriented fashion, with
	support for defining sums and products of operators with simple arithmetic
	operations.

	The potential part of the Hamiltonian operator is implemented with direct
	pointwise multiplication of the wave function with precomputed values. A
	few common potential types are provided, and implementing new ones is as
	easy as providing a C++ routine which gives the values of the potential as
	a function of position. There is also rudimentary support for adding
	arbitrary types of random noise to the potentials.

	In the case of periodic boundary conditions with no magnetic field the
	kinetic energy part of the Hamiltonian is implemented by simply expanding
	the wave function in plane waves via a discrete Fourier transform,
	multiplying with $\vec{k}^2/2$, where $\vec{k}$ is the wave vector, and
	returning to the position basis via an inverse discrete Fourier transform.
	External magnetic field only shifts the eigenvalues of the momentum
	operator with the vector potential~$\vec{A}$, so in the case of nonzero
	magnetic fields, the states simply need to be multiplied with
	$(\vec{k}+\vec{A})^2/2$. As discussed in Sec.~\ref{s:magnetic}, the magnetic
	field is assumed to be homogeneous and parallel to the calculation plane,
	and for numerical efficiency all wave functions and operators are expressed
	in the linear gauge $\vec{A} = (-By, 0, 0)$. For Dirichlet boundary conditions,
	states are expanded in eigenstates of the particle in a box (via a discrete
	sine transform). The kinetic energy operator with no magnetic field is
	again a simple multiplication in the sine function basis, but as discussed
	in Sec.~\ref{s:dirichlet}, the case of nonzero magnetic field is more
	complicated. In this case the operator is split into two parts,
	$(-\i\nabla+\vec{A})^2/2 = (-\nabla^2 + \vec{A}^2)/2
	-\i\vec{A}\cdot\nabla$, so that the first part is a simple multiplication
	in the sine function basis, and the second one turns the sine
	functions into cosines multiplied by a suitable factor, i.e., it is a sine
	transform followed by a multiplication and an inverse \emph{cosine}
	transform.

	The exponentiated operators $\exp(-\eps V)$ and $\exp(-\eps T)$ required
	for imaginary time propagation are implemented using Fourier or sine
	transforms in a similar way as the original potential and kinetic energy
	operators. The exponentiated potential operator is still a pointwise
	multiplication in the position basis, and as discussed in Sec.~\ref{s:magnetic}
	the exponentiated kinetic energy operator can be factorized into parts that
	can be implemented with discrete Fourier transforms and pointwise
	multiplications. In both cases the multiplication arrays are precomputed
	and only recalculated when the time step~$\eps$ is changed. The full
	imaginary time propagation operator is then built from the two operators by
	operator arithmetic as specified by the Chin factorization~\cite{chin} given in
	Eq.~\eqref{eq:chin}, up to the order specified by the user. The
	resulting chain of operator sums and products is simplified when possible
	by absorbing constant prefactors into the operators themselves and
	combining adjacent multiplications.

	The orthonormalization of wave functions is implemented using the subspace orthonormalization
	method described in Sec.~\ref{s:ortho}. The inner products and the
	diagonalization of the overlap matrix are simply delegated to external
	linear algebra routines. For the linear combination two alternative
	algorithms are provided: the default one in which the large matrix
	product~\eqref{eq:lincomb} is split into matrix-vector products in order to
	use as little extra memory as possible, and one where the product is
	calculated directly, requiring an extra copy of the wave functions. The
	latter algorithm may be faster in some cases since it makes maximal use of
	optimized external routines, but this is offset by the roughly double
	memory requirement which makes a huge difference for large systems.

	The ITP cycle is started with random noise as the initial wave functions.
	This helps to ensure that no eigenstate of the Hamiltonian is missing from
	the expansions of the initial states due to accidental orthogonality, as
	discussed in Sec.~\ref{s:itp}. The desired number of states to be converged is
	provided by the user, as is the total number of states to be included in the
	computation. If the latter is missing, the program adds 25\% to the number
	required to converge. This default value was empirically determined to
	provide a good convergence speed. The initial imaginary time step~$\eps$ is
	also provided by the user, and after all the required states have converged
	with respect to the time step size, it is decreased by dividing by a
	user-provided constant. It is also possible to fine-tune the convergence by
	explicitly listing all time step values that are to be used. Each time
	convergence with respect to the time step is found, states are also tested
	against the criteria of final convergence, and if it is fulfilled by all the
	required states the computation ends. The criteria for convergence are
	also provided by the user. By default the program uses the standard
	deviation with respect to the Hamiltonian, but other criteria listed
	in Sec.~\ref{s:convergence} have also been implemented.

\subsection{Data file output}

	In addition to the textual output given by the command-line interface,
	\texttt{itp2d} saves its results and parameters as portable HDF5 data
	files. All data coming from a single simulation run are saved in a single
	data file. The user can specify whether, in addition to the parameters
	given to the simulation, only final energies (along with their error
	estimates) are saved, or also the final wave functions, or even
	intermediate wave functions after each iteration. Using a common (as
	opposed to application-specific) data file format has several advantages.
	First of all, the data can be imported easily to common data analysis
	software, and accessing the data is easy: the HDF5 format presents the data
	as a directory of data sets, with descriptive names for each set. With HDF5
	even complicated, multidimensional data can be saved without trouble and
	without complicating later data access.

\subsection{Parallelization}

	Many parts of the ITP computation are suitable for shared-memory
	parallelization. The most trivial case is the actual propagation step,
	where each state is operated on by the imaginary time propagation operator
	independent of each other. In a similar way, the task calculating the
	energy and standard deviation of energy for each state is trivially split
	to several, independent processing threads. The orthonormalization step is
	not as easily parallelized, but most of the work can be distributed by
	calculating the inner products for the overlap matrix in parallel, and
	splitting the matrix product of Eq.~\eqref{eq:lincomb} into
	matrix-vector products which are executed in parallel.

	In all above cases our implementation uses high-level OpenMP instructions
	for parallelization, making the parallelized code simple and readable. It
	should be noted that due to the large amount of data that needs to be
	passed to the execution threads, especially for parallelizing the
	orthonormalization step, ITP works best with shared-memory parallelization,
	i.e., several execution cores accessing the same physical memory. When
	using the program in large cluster computers special care should be taken
	to ensure that there is no unnecessary memory access across slow network
	links.

\subsection{External routines}

	To avoid needless reimplementation, most low-level numerical
	operations used in~\texttt{itp2d} are delegated to external routines. This
	also allows the user to use routines heavily optimized to the current
	hardware. All linear algebra routines are accessed via standard LAPACK and
	CBLAS interfaces. Our program has been tested with the portable ATLAS
	implementation, the MKL library from Intel, and the ACML library from AMD.
	Discrete Fourier, sine and cosine transformations are computed using the
	heavily optimized library FFTW3~\cite{fftw}. Intel's MKL library provides a
	FFTW3-compliant interface, but using MKL for the transformations has not
	been tested with our program.

\subsection{Provided unit tests}

	Our program is distributed with a comprehensive unit test suite, implemented
	with the Google C++ Testing Framework. The unit tests cover several low-level
	details, such as the accuracy of external Fourier transform routines and
	the internal logic of several arithmetic operations, and high-level features,
	such as running ITP simulations using potentials with known analytic
	eigenstates and comparing the results. It is advisable to always run this unit
	tests suite before important calculations to protect against unforeseen errors.

\subsection{Open source}

	We release~\texttt{itp2d} under an open-source license with the intention
	that it will foster wide use and future development of the code. Users are
	encouraged to improve and extend the code and share their changes with other
	users of~\texttt{itp2d}. More information about getting involved can be
	found in the \texttt{README} file distributed with \texttt{itp2d}.

\section{Numerical tests}

\subsection{The harmonic oscillator}

	The harmonic oscillator is an example of a system with a known analytic
	solution with or without a magnetic field. The harmonic oscillator with
	potential~$V(\vec{r}) = \tfrac{1}{2}r^2$ is also the default potential used in
	\texttt{itp2d}, so simply running the program with no additional command line
	parameters will produce the first few eigenstates of the harmonic oscillator.
	The energy levels of the above potential will follow the Fock-Darwin spectrum
	\begin{equation}
		\label{eq:fd}
		E_{nl}(B) = (2n + |l| + 1)\sqrt{1 + \tfrac{1}{4}B^2} - \tfrac{1}{2}lB,
	\end{equation}
	with $n = 0, 1, 2, \dotsc$, $l = 0, \pm1, \pm2, \dotsc$.

	\begin{table}
		\centering
		\caption{Calculated energy levels of a harmonic oscillator potential
		$V(\vec{r}) = \tfrac{1}{2}r^2$ with magnetic field strength~$B=1$. The
		table also shows the standard deviation of energy~$\sigma_H$, the exact
		value of energy, and the actual error of the result. All results are
		from a single simulation with 5000 states. The final convergence
		criteria was that the first 4000 states have $\sigma_H/E < 10^{-3}$. As
		is evident from the results, the lower energy states converge to a lot
		higher accuracy.}
		\begin{tabular}{rr@{.}llr@{.}ll}
			\toprule
			\#&
			\multicolumn{2}{c}{$E$} &
			\multicolumn{1}{c}{$\sigma_H$} &
			\multicolumn{2}{c}{$E_\text{exact}$} &
			\multicolumn{1}{c}{$|E-E_\text{exact}|$}\\
			\midrule
			0	&  1&118033988749895& $6\times10^{-8}$	&  1&118033988749895& $< 10^{-15}$\\
			10	&  4&5901699437497  & $2\times10^{-11}$	&  4&590169943749475& $3\times10^{-13}$\\
			100	& 14&152475842500   & $4\times10^{-11}$	& 14&152475842498529& $3\times10^{-12}$\\
			400	& 28&311529493753   & $5\times10^{-11}$	& 28&31152949374527	& $9\times10^{-12}$\\
			1000& 44&74922359501    & $8\times10^{-12}$	& 44&74922359499622	& $3\times10^{-11}$\\
			2000& 63&25580140378    & $6\times10^{-11}$	& 63&25580140374443	& $4\times10^{-11}$\\
			4000& 89&4495           & $6\times10^{-2}$	& 89&44929690873981	& $3\times10^{-4}$\\
			\bottomrule
		\end{tabular}
		\label{tab:fd}
	\end{table}

	Table~\ref{tab:fd} collects some eigenenergies computed for the harmonic
	oscillator by~\texttt{itp2d} from a simulation with 4000 states required to
	converge (5000 states in total) and magnetic field strength $B=1$. The
	final convergence criteria used was a relative standard deviation of energy
	$\sigma_H/E$ of less than $10^{-3}$. As is seen in the table, the accuracy
	of computed energies is very good up to highly excited states, and in most
	cases~$\sigma_H$ gives a good upper bound estimate of the actual error in
	the result. For state 1000 the standard deviation~$\sigma_H$ is less than
	the actual error. In general, very small values of~$\sigma_H$ are not
	reliable due to discretization errors. This simulation required~5
	iterations of ITP starting with time step $\eps=0.1$. The simulation used
	a~500 by~500 grid and 12th order operator splitting.

\subsection{High energy eigenstates of a particle in a box with a magnetic field}

	The particle in a box, i.e., a potential that is zero inside a rectangular
	box and infinite elsewhere, is another example of a potential with a known
	energy spectrum -- except for the case with nonzero magnetic field. With a
	magnetic field the system is no longer trivial, and no analytic solution is
	known. Another interesting feature of this system is that the corresponding
	classical system shows chaotic behavior~\cite{berglund,esasquare}, making the particle in a
	box with magnetic field an interesting testbed for quantum chaos studies.

\begin{figure}[htb]
		\centering
		\begin{tabular}{c@{\hspace{0.01\textwidth}}c@{\hspace{0.01\textwidth}}c}
		\includegraphics[width=0.30\textwidth]{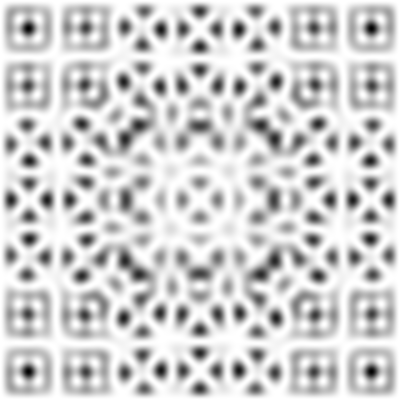}&
		\includegraphics[width=0.30\textwidth]{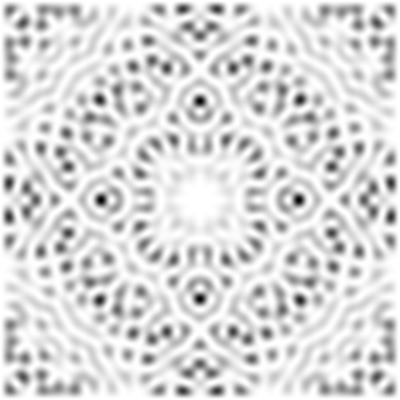}&
		\includegraphics[width=0.30\textwidth]{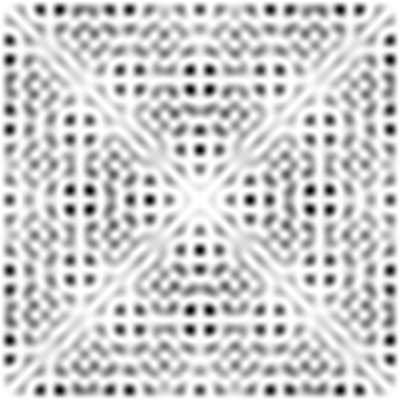}\\
		365th & 546th & 622th\\[0.75em]
		\includegraphics[width=0.30\textwidth]{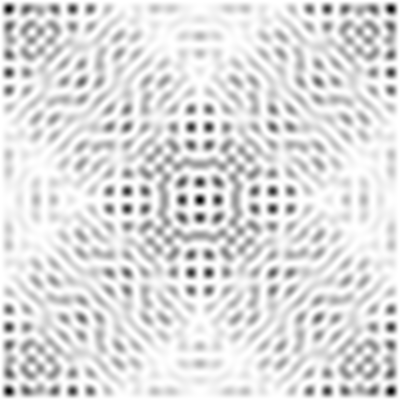}&
		\includegraphics[width=0.30\textwidth]{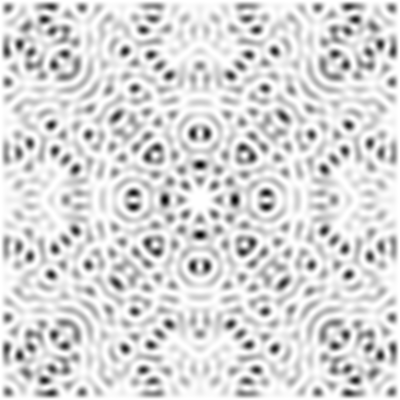}&
		\includegraphics[width=0.30\textwidth]{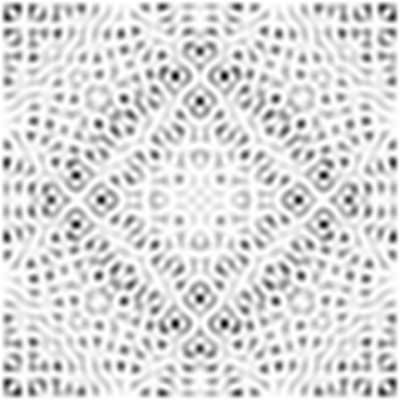}\\
		750th & 890th & 975th
		\end{tabular}
		\caption{Density plots of a few collected eigenstates of the particle
		in a box with a magnetic field. The eigenstates were calculated by
		\texttt{itp2d} for a $\pi$ by $\pi$ box with magnetic field
		strength~$B=1$.}
		\label{fig:carpets}
	\end{figure}

	Since the system combines Dirichlet boundary conditions and a magnetic
	field, it is subject to the problems discussed in~Sec.~\ref{s:dirichlet}, i.e.,
	slower convergence and inaccurate calculation of~$\sigma_H$. However, in
	order to illustrate that these problems do not prevent the study of this
	system with \texttt{itp2d}, we demonstrate that we have calculated a
	thousand eigenstates of this system. However, since no analytic expression
	of the energy is known, and since we are not aware of any other program
	that could compute this many eigenstates of this particular system,
	assessing the accuracy of the calculation is difficult. The ITP calculation
	still converges, and the wave functions show no sign of numerical
	error. Due to the complicated interplay of the magnetic field and the
	``hard'' potential walls, the eigenstate wave functions have a very
	intricate form, as shown in Figure~\ref{fig:carpets}. The
	wave functions of a square billiard in magnetic field have been reported
	previously only for the first few eigenstates~\cite{square, square2}.

	In our future studies we will focus on the chaotic properties of the
	present and other billiard systems in magnetic fields by examining the
	spectral properties in detail. The itp2d code is a versatile tool for that
	purpose.

\subsection{Benchmark results}
\label{s:benchmarks}

\begin{figure}[htp]
		\begin{center}
			\includegraphics[width=0.8\textwidth]{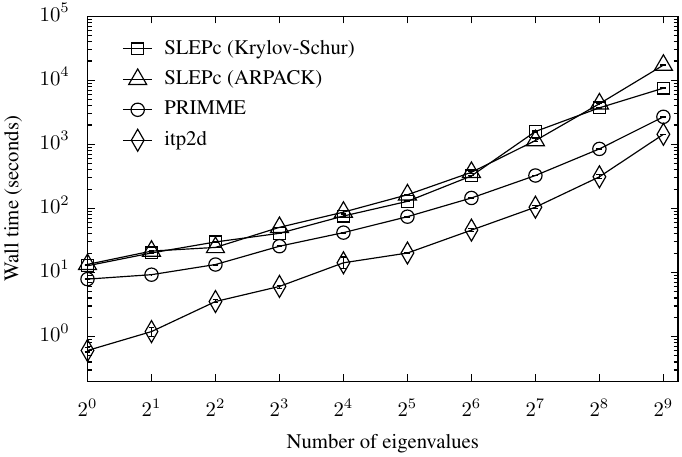}
		\end{center}
		\caption{Elapsed wall time in an eigenstate computation as a function
			of the number of eigenstates. The results are shown for four
			different solvers: SLEPc using its default solver algorithm
			(Krylov-Schur), SLEPc interface for ARPACK, PRIMME, which
			implements several solver algorithms and chooses the optimal one
			dynamically, and \texttt{itp2d}. All computations were repeated
			three times, and the average wall time was used for making the
			figure. The error bars on the data points show the minimum and
			maximum wall time value, respectively. The same Hamiltonian
			operator implementation, the same grid size and the same
			convergence criterion were used for all computations. The
			computations were done on a single, dedicated workstation.}
		\label{fig:bench}
	\end{figure}

	To assess the numerical efficiency of our program, we have benchmarked
	\texttt{itp2d} against publicly available general-purpose eigensolvers. The
	solvers used in our test were PRIMME~\cite{primme} (version 1.1) and
	SLEPc~\cite{slepc} (version 3.3-p2), the latter also functioning as an
	interface for ARPACK (arpack-ng version 3.1.2). Our benchmark consisted of
	solving an increasing number of eigenstates of a quartic oscillator
	potential in zero magnetic field, and measuring the elapsed wall time. The
	computations were run without parallelization. The results of this test are
	shown in Figure~\ref{fig:bench}.

	As with any benchmark, the results need to be interpreted with care. It is
	very difficult to compare the performance of different programs and
	different algorithms completely fairly. Besides \texttt{itp2d}, the tested
	eigensolvers only use the Hamiltonian of the system to compute the
	eigenstates. In this test all eigensolvers use the Hamiltonian implemented
	in \texttt{itp2d}, which provides \texttt{itp2d} with an advantage. The
	other solvers compute matrix-vector products of the Hamiltonian and a state
	vector out-of-place (i.e., without overwriting the original state), whereas
	the Hamiltonian implementation of \texttt{itp2d} is in-place. This incurs
	an overhead, since each matrix-vector product requires the vector to be
	copied. This overhead was measured to be small (a few percent of total
	runtime), and it was subsequently subtracted from the results show in
	Figure~\ref{fig:bench}. Some advantage still remains from the fact that
	in-place Fourier transforms computed with FFTW3 are somewhat slower than
	their out-of-place variants~\cite{fftw}. Another issue which complicates
	the interpretation of this simple benchmark is that all the tested programs
	and algorithms have several parameters, and truly optimal performance would
	require fine-tuning these parameters for each system and problem size. For
	example, different preconditioning strategies were not tested for any of
	the contestants.

	As a conclusion, even though our simple benchmark does not capture the
	whole truth regarding the performance of \texttt{itp2d} compared to other
	programs, we are confident that \texttt{itp2d} performs on a level which is
	comparable to other eigensolver implementations.

\section{Summary}

	The program we have presented, \texttt{itp2d}, is a modern 
	implementation of the imaginary time propagation algorithm for solving the
	single-particle, time-independent Schr\"odinger equation in two dimensions.
	Its strengths include a clear, object oriented design, and the ability to
	include a strong, homogeneous magnetic field. It released with the aim of providing researchers with a flexible
	and extensible code package for solving the eigenstates and energy spectra
	of two-dimensional quantum systems.

	As immediate applications, we find appealing possibilities in the field of
	quantum chaos in terms of spectral statistics. Furthermore, it is
	straightforward to combine itp2d with real-space electronic-structure
	methods based on density-functional theory, e.g., into the Octopus code
	package~\cite{octopus}.

\section{Acknowledgements}

	This work was supported by the Finnish Cultural Foundation, University of
	Jyv\"askyl\"a and the Academy of Finland. We are grateful to CSC -- the
	Finnish IT Center for Science -- for providing computational resources. We
	wish to thank P.~Heliste for assistance in testing~\texttt{itp2d}, and
	Prof.~A.~Stathopoulos for useful discussions.

\section*{References}

\bibliographystyle{elsarticle-num-names}
\bibliography{itp2d-paper}

\end{document}